\documentclass[smallextended]{svjour3}       
\smartqed  
\usepackage{graphicx}
\usepackage{times}
\usepackage{ulem}
\usepackage{array}
\usepackage{amsmath}
\usepackage{color}

\journalname{Quantum Information Processing}
\begin{document}

\title{Quantum walks on a circle with optomechanical systems
}


\author{Jalil Khatibi Moqadam \and
        Renato Portugal \and
        Marcos Cesar de Oliveira
}


\institute{Jalil Khatibi Moqadam \at
              Laborat\'orio Nacional de Computa\c c\~ao Cient\'\i fica (LNCC),
              Petr\'opolis, RJ, Brazil\\
              \email{jalilkhm@hotmail.com}\\
              \emph{Present address:}
              Instituto de F\'isica ``Gleb Wataghin'', Universidade Estadual de Campinas,
              Campinas, SP, Brazil\\
           \and
           Renato Portugal \at
              Laborat\'orio Nacional de Computa\c c\~ao Cient\'\i fica (LNCC),
              Petr\'opolis, RJ, Brazil\\
              \email{renato.portugal.d.sc@gmail.com}
           \and
           Marcos Cesar de Oliveira \at
              Instituto de F\'isica ``Gleb Wataghin'', Universidade Estadual de Campinas,
              Campinas, SP, Brazil\\
              \email{marcos@ifi.unicamp.br}
}

\date{Received: date / Accepted: date}

\maketitle

\begin{abstract}
We propose an implementation of a quantum walk on a circle on an optomechanical system by encoding the walker on the
phase space of a radiation field and the coin on a two-level state of a mechanical resonator. The dynamics of the system
is  obtained by applying Suzuki-Trotter decomposition. We numerically show that the system displays typical behaviors of
quantum walks, namely, the probability distribution evolves ballistically and the standard deviation of the phase
distribution is linearly proportional to the number of steps. We also analyze the effects of decoherence by using the
phase damping channel on the coin space, showing the possibility to implement the quantum walk with present day technology.

\end{abstract}

\section{Introduction}
Discrete time quantum walks are the counterpart of the classical random walks when the coin used to direct the walker steps
can assume a superposition of its possible states \cite{Aharonov1993}.  This ability to exist in a superposition leads to a
highly entangled joint state for the coin and walker over time \cite{Vieira2013}, resulting in a clear speedup. Therefore it
is not surprising that similarly to its classical version, quantum walks are connected to the success of search algorithms,
depending on the pertinent geometry \cite{Childs2003,Childs2004}. 
This success has promoted a broad investigation over the last years, from computational (algorithmic) point of
view \cite{Portugal2013} or in order to efficiently implement it physically \cite{Manouchehri2014}. Actual implementations
were reported through experiments on
optical \cite{Bouwmeester1999,SoutoRibeiro2008,Broome2010,Zhang2010,Peruzzo2010,Schreiber2010,Schreiber2011,Goyal2013}
and atomic systems
\cite{Xue2009,Karski2009,Zahringer2010,Schmitz2009}. 

Particularly interesting from the quantum optics perspective is the quantum walk on a circle since it can be implemented on
a large number of systems, with a variable amount of control. The general idea is that the walker steps are encoded as phase
increments in the phase space of a harmonic oscillator, whose signal is conditioned to an auxiliary quantum coin. Some
examples of systems explored include micromasers \cite{Aharonov1993}, optical lattices \cite{Dur2002}, ion
traps \cite{Travaglione2002}, cavity quantum electrodynamics \cite{sanders2003quantum,xue2008quantum}, as well as
superconducting circuit quantum electrodynamics \cite{Xue2008} and
an ensemble of nitrogen-vacancy centers in diamond~\cite{hardal2013discrete}.
A natural system where the quantum walk on a circle can
be implemented though is on the emerging field of quantum optomechanics, where quantum mechanical  modes can be coupled to
quantum light modes \cite{aspelmeyer2013cavity}. Given the degree of control achieved  and the ability to reach an enormous
variety of strong coupling in those systems \cite{Poot201227}, a particular advantage over the previous experimental
realizations would be the possibility to implement a quantum walk on a large number of time steps. It is therefore
expected that they might be employed in the near future for implementation of a quantum walk, be it on the optical
or the microwave regime. 

\begin{figure}[htbp]
\begin{center}
\includegraphics[width=.45\textwidth]{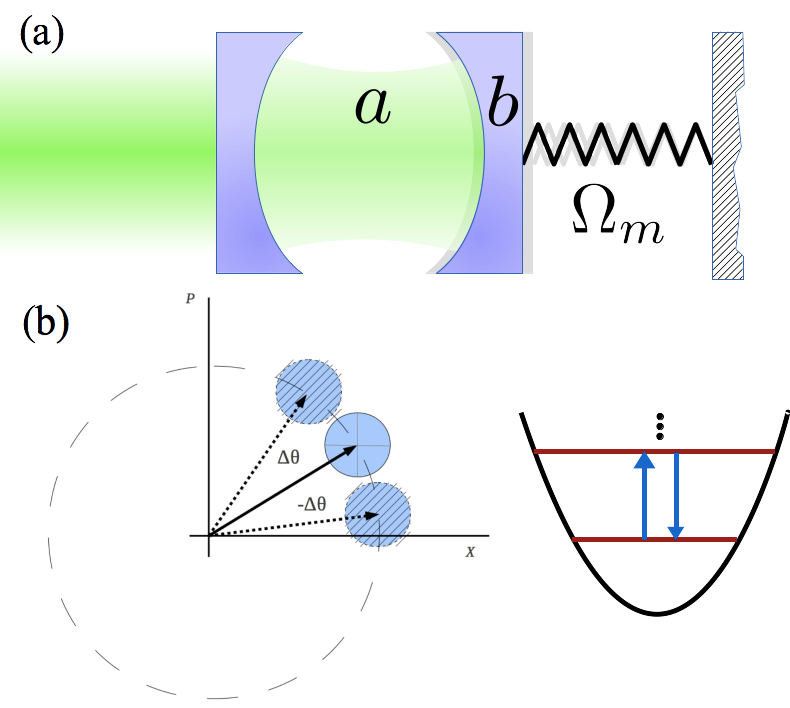}
\end{center}
\caption{(Color Online) Quantum walk implemented with a typical quantum optomechanical system (a) where a radiation
mode $a$ is coupled to a mechanical mode of frequency $\Omega_m$.  (b) The optical field described by a coherent state
has its phase shifted by $ \Delta\theta$ or $- \Delta\theta$ conditioned on the state of the quantum coin encoded in
the two lowest levels of the mechanical mode.}\label{fig1}
\end{figure}

Here we propose an implementation of a quantum walk on a circle in an optomechanical (or electromechanical) system.
The proposed system is an optomechanical resonator, which is composed of an optical cavity coupled to a mechanical
resonator as presented in Fig.~\ref{fig1}(a). This kind of system has been explored in many physical
experiments \cite{aspelmeyer2013cavity}, and the quantum regime has been recently
reached \cite{o2010quantum,Chan2011,Teufel2009,Poot201227}. We assume a dispersive interaction between the mechanical
mode and the radiation, which that can in principle be achieved in both optical and microwave regimes due to radiation
pressure \cite{Teufel2009}.
This interaction provides a way to implement the operations employed in quantum walks. The position of a coherent state
in the phase space for the radiation mode is used to encode the position of the walker, while the mechanical mode plays
the role of a quantum coin. 

We investigate the situation for a two-sided coin by considering low excitation of the mechanical resonator, where
it can be appropriately described by a two-level system. The shift operator
is implemented via the coupling to the mechanical resonator. By adjusting the frequency, its evolution is driven
by a Hadamard-like operator, necessary in the evolution.  By performing numerical simulations, we show that the
dynamics of the system present the main signature of the quantum walk evolution, namely, the ballistic spread of
the phase probability distribution. The transition from quantum to classical walk due to noise effects on the
coin is investigated by introducing decoherence via the phase damping channel on the coin space.

The paper is structured as follows. In Sec.~\ref{sec:QW} we derive the dynamics of the systems and describe 
how the coined quantum walk can be implemented.
In Sec.~\ref{sec:decoherence} we analyze the decoherence by using the phase damping channel affecting the coin.
Finally, the summary and discussion are presented in Sec.~\ref{sec:discussion}.

\section{Quantum Walk in Phase Space of an Optical Resonator}\label{sec:QW}
The optomechanical Hamiltonian in a frame rotating with the drive frequency is
given by~\cite{aspelmeyer2013cavity}
\begin{equation}
 \label{eq:standard_om_hamiltonian}
 \mathcal{H} =- \hbar \Delta a^{\dagger}a + \hbar \Omega_m b^{\dagger}b
                - \hbar g_0 a^{\dagger}a (b^{\dagger} + b) + \hbar \mathcal{E} (a^{\dagger} + a),
\end{equation}
where $a^{\dagger}$ ($a$), $b^{\dagger}$ ($b$) are the creation (annihilation) operators for the optical and
mechanical resonators respectively, $\Delta$ is the drive field detuning with respect to the cavity resonance mode,
$\Omega_m$ is the mechanical resonator frequency, $g_0$ is the photon-phonon interaction strength, which is given due
to radiation pressure force, and
$\mathcal{E}$ is the drive amplitude.\footnote{Usually in Hamiltonian (\ref{eq:standard_om_hamiltonian}) a linearization
procedure is performed by assuming an undepleted cavity field, where the field operator is given by $a\approx \alpha+\delta a$,
being $\alpha$ a coherent amplitude and $\delta a$ a quantum fluctuation.  Had we considered this linearization we would have
ended up with the interaction term $-\hbar g_0 (\alpha \delta a^{\dagger}+\alpha^* \delta a) (b^{\dagger} + b)$, which turns
out to lead to a quantum walk on a line given the displacement operator for the field $(\delta a^{\dagger}+ \delta a)$
for $\alpha$ real. In this paper, we do not consider this procedure.}

In a stationary regime the driving $\mathcal{E}$ competes with dissipative effects (described by an amplitude damping
channel at $T=0$) forcing the optical field to be in a coherent state, a state with a well defined
phase [see Fig.~\ref{fig1}(b)].
The coupling considered induces a frequency shift on the optical field conditioned on the displacement of the mechanical
resonator. Since the optical field is in a coherent state, the frequency shift translates into a wandering of the optical
field phase conditioned on the excitations of the mechanical resonator. As will be shown ahead, it makes sense to analyze the
action of
the interaction Hamiltonian apart from the other terms in (\ref{eq:standard_om_hamiltonian}).
If the optical mode is in a coherent state, then the interaction Hamiltonian 
induces a phase shift on the coherent state conditioned on the displacement of the mechanical resonator.
This feature suggests that the coupling
should be explored as the shift operator in the standard quantum walk. This is best appreciated if we consider the
situation where the mechanical resonator is cooled down to its ground
state \cite{o2010quantum,Chan2011,Teufel2009,Poot201227} and from there can be coherently promoted to the first excited
state, such that, the average number of phonons $\langle b^{\dagger}b \rangle$ is smaller than 1. The mechanical resonator can
therefore be approximately described by only its two lowest energy levels and its creation and annihilation operators
are replaced, respectively, by the spin raising and
lowering operators $\sigma_{+} = | e \rangle \langle g |$ and $\sigma_{-} = | g \rangle \langle e |$, where $| g \rangle$
and $| e \rangle$ are the ground and the first exited states of the mechanical resonator, constituting a two-sided
quantum coin.
Relations $\sigma_{+}\sigma_{-}=(1/2)(I + \sigma_z)$ and $\sigma_{+}+\sigma_{-}=\sigma_x$ can then be used
and, by performing a
basis change to the
rotated frame $| \pm \rangle = (1/\sqrt{2})(| e \rangle \pm | g \rangle )$, Pauli matrices $\sigma_x$ and $\sigma_z$ are
replaced by  $\tilde{\sigma}_z$ and  $\tilde{\sigma}_x$, respectively. Neglecting the constant energy, the
Hamiltonian~(\ref{eq:standard_om_hamiltonian}) can be rewritten as
\begin{equation}
 \label{eq:standard_om_hamiltonian_two_level_rotated}
 \mathcal{H} = - \hbar \Delta a^{\dagger}a
               - \hbar g_0 a^{\dagger}a \tilde{\sigma}_z
               + \hbar \mathcal{E} (a^{\dagger} + a)
               + \frac{1}{2}\hbar \Omega_m \tilde{\sigma}_x.
\end{equation}

Using the Suzuki-Trotter formula \cite{suzuki1985decomposition} we can express the evolution operator, given the 
Hamiltonian~(\ref{eq:standard_om_hamiltonian_two_level_rotated})  as
\begin{eqnarray}
 \label{eq:Suzuki_Trotter}
 \mathcal{U}(T) &=& e^{ -i \mathcal{H} T / \hbar} \nonumber\\
                &=& \lim_{n \to \infty} \{
                                      e^{ iT \Delta a^{\dagger}a / n }\;.\;  e^{ iT g_0 a^{\dagger}a \tilde{\sigma}_z / n } \;
                                      .\;e^{-iT \mathcal{E} (a^{\dagger} + a) / n }\;.\;
                                      e^{-iT \Omega_m \tilde{\sigma}_x /(2n) }
                                      \}^n, 
\end{eqnarray}
where $T$ is the total time evolution and $n$ is the number of total time partitions.
Generally for any set of skew-Hermitian operators $\{H_j;j=1,...,p\}$ in a Banach space, the first order Suzuki-Trotter
approximation error $\epsilon$ is bounded as~\cite{suzuki1985decomposition,suzuki1976generalized}
\begin{equation}
 \label{eq:suzuki_theorem}
 \biggl \| \exp \biggl( \sum_{j=1}^{p} H_j \biggr) -
 \biggl[ \prod_{j=1}^{p} \exp{\left(\frac{1}{n} H_j\right)} \biggr]^n \biggr\| 
 \; \leq \; \frac{1}{2n} \sum_{j>k} \| [ H_j,H_k ] \|.
\end{equation}
Actually, the limit in Eq.~(\ref{eq:Suzuki_Trotter}) is a result of Eq.~(\ref{eq:suzuki_theorem}) provided that the
operators are bounded. It must be mentioned that the annihilation and creation operators {\color{black}$a$ and $a^{\dagger}$}
are not bounded in general. However, they remain bounded when restricted to operate on the coherent states with bounded
displacement of the vacuum state.

The upper bound error corresponding to  Eq.~(\ref{eq:suzuki_theorem}) for a
finite $n$ is then
calculated as
\begin{equation}
 \epsilon \; \leq \;\; \frac{T^2}{2n} \biggl( \Delta \mathcal{E} \| a^{\dagger} - a \| +
                                                \Omega_m g_0 \| \tilde{\sigma}_y a^{\dagger}a \| 
                                              + g_0 \mathcal{E} \| \tilde{\sigma}_z ( a^{\dagger} - a ) \| \biggr).
\end{equation}
Using the operator norm
\[ \| H_j \| := \sup_{\| |\psi\rangle\| = 1 } \| H_j |\psi\rangle \|, \]
with $|\psi\rangle = |q\rangle |\alpha\rangle$, where $|q\rangle$ is the state of the qubit and $|\alpha\rangle$ is the
coherent state of the optical field, the error can be written as
\begin{align}
 \label{eq:error}
 \epsilon \; \leq \;\; \frac{T^2}{2n} \biggl[
                        \varepsilon &(\Delta+g_0) \left( |\alpha_{\mathrm{max}}|
                                       + \sqrt{1+|\alpha_{\mathrm{max}}|^2} \right) \biggr. \\ \nonumber
                                    & \biggl. + \; \Omega_m g_0 |\alpha_{\mathrm{max}}|\sqrt{1+|\alpha_{\mathrm{max}}|^2}
                                    \;\biggr],
\end{align}
where $\alpha_{\mathrm{max}}$ has the largest modulus over the whole desired (bounded) region of the phase space.
Therefore, by taking $n$ sufficiently large the error can be made as small as required.

\subsection{Quantum walk dynamics}\label{sec:dynamics}

Taking $n$ large enough as discussed above and choosing the mechanical resonator frequency so that
\begin{equation}
 \label{eq:coin_cond}
 \tau \equiv\frac{T}{n} = \frac{\pi }{ 2 \Omega_m}, 
\end{equation}
the approximate evolution of the system is given by
\begin{equation}
 \label{eq:Suzuki_Trotter_qw}
 \mathcal{U}(T=n\tau) \approx U^n,
\end{equation}
where
\begin{equation}\label{eq:U}
U = e ^{ i \tau \Delta a^{\dagger}a }
                     e ^{ i \tau g_0 a^{\dagger}a \tilde{\sigma}_z }
                     e ^{ - i \tau \mathcal{E} (a^{\dagger} + a) }
                     e ^{ - i \frac{\pi}{4} \tilde{\sigma}_x }.
\end{equation}
The evolution (\ref{eq:Suzuki_Trotter_qw})
corresponds to a $n$-step discrete-time quantum walk when $\tau \mathcal{E}$ is small. The state of the quantum
walk is the product of the mechanical resonator two-level state with the coherent state of the optical field.
The coherent state is conditionally shifted depending on the state of the qubit.
Specifically, the last exponential of unitary evolution~(\ref{eq:U}) can generate a Hadamard-like
transformation for the qubit since
\begin{eqnarray}
 \label{eq:hadamard}
 e^{-i \frac{\pi}{4} \tilde{\sigma}_x }  &=&
 \frac{1}{\sqrt{2}} \begin{pmatrix}
                    \;1\ \;-i\\
                     -i\ \;\;\; 1\\
                    \end{pmatrix}\nonumber\\
                    &&=\frac{1}{\sqrt{2}} \left(
\begin{array}{cc}
1  &  0  \\
0 &   -i 
\end{array}
\right)                    \left(
\begin{array}{cc}
1  &  1  \\
1 &   -1 
\end{array}
\right)\left(
\begin{array}{cc}
1  &  0  \\
0 &   -i 
\end{array}
\right).\end{eqnarray}
This term plays the role of the coin operator.

The coherent state of the optical resonator
\begin{equation}
\label{eq:coherent_state}
  | \alpha \rangle = e^{-\frac{1}{2}|\alpha|^2} \sum_{j=0}^{\infty} \frac{\alpha^j}{\sqrt{j!}} | j \rangle,
\end{equation}
where $| j \rangle$ are the number states of the resonator, is rotated in the phase space,
as in  Fig.~\ref{fig1}(b)
for $\Delta\theta= \tau g_0$ under the
application of the second exponential of unitary evolution~(\ref{eq:U})
\begin{align}
 \label{shift_op}
             e^{ i \tau g_0 a^{\dagger} a \tilde{\sigma}_z } | \alpha \rangle | \pm \rangle
                                    = | \alpha e^{\pm i \tau g_0 } \rangle | \pm \rangle,
\end{align}
conditioned on the coin state $|\pm\rangle$.
This term plays the role of the shift operator \cite{Travaglione2002,Xue2008}.
The evolution driven by the interaction Hamiltonian leads to an entangled state
when the mechanical resonator is in a superposition of the ground and the first excited states.

The third exponential of unitary evolution~(\ref{eq:U}) disturbs the quantum walk dynamics, but it can be made small enough
in order to keep the quantum walk characteristics. Finally, the first exponential is a free rotation that is not related with
the
shift operator (second exponential). Its effect is an overall phase shift. The effects of the first and third terms are also
discussed later.

It is now possible to find the dynamics of the system during the time $T=n\tau$. Starting with the initial state
\begin{equation}
\label{eq:asym_ini}
| \psi_0 \rangle = | + \rangle | \alpha_0 \rangle,
\end{equation}
after the first time step the system evolves to $$|\psi_1 \rangle = U | + \rangle | \alpha_0 \rangle,$$
which can be eventually simplified to
\begin{equation}
 |\psi_1 \rangle = c_1^{(1)} | + \rangle | \alpha_{1}^{(1)} \rangle
                                           + c_2^{(1)} | - \rangle | \alpha_{2}^{(1)} \rangle,
\end{equation}
where
\begin{equation}
 \label{eq:alpha_1}
 \left\{
  \begin{array}{ll}
   \alpha_{1}^{(1)} = ( \alpha_0 - i \tau \mathcal{E} ) e^{ i \tau ( \Delta + g_0  )}, \\
   \alpha_{2}^{(1)} = ( \alpha_0 - i \tau \mathcal{E} ) e^{ i \tau ( \Delta - g_0  )},
  \end{array}
 \right.
\end{equation}
and
\begin{equation}
 \label{eq:coeff_1}
 \left\{
  \begin{array}{ll}
   c_{1}^{(1)} = \frac{1}{\sqrt{2}} e^{ -i  \tau \mathcal{E}  \mathrm{Re} ( \alpha_0 ) },\\
   c_{2}^{(1)} = \frac{-i}{\sqrt{2}} e^{-i  \tau \mathcal{E}  \mathrm{Re} ( \alpha_0 ) }.
  \end{array}
 \right.
\end{equation}

In the same way it is possible to calculate the state of the system in the next time steps.
In each time step the number of terms is twice the number of terms in the previous state.
Associating the odd subindices to the terms with $| + \rangle$ and the even subindices to the terms with $| - \rangle$
in all time steps, the state of the system at step $l$ can be written as
\begin{equation}
 \label{eq:state_at_step_l}
 | \psi_l \rangle = \sum_{k=1}^{2^{l-1}}
       \left\{
             c_{2k-1}^{(l)} | + \rangle | \alpha_{2k-1}^{(l)} \rangle +
             c_{2k}^{(l)} | - \rangle | \alpha_{2k}^{(l)} \rangle
       \right\},
\end{equation}
where $1 \leq l \leq n$. Recursively, the complex numbers $\alpha$ are obtained as
\begin{equation}
 \label{eq:alpha's}
 \left\{
  \begin{array}{ll}
   \alpha_{2k-1}^{(l)} &= [ \alpha_k^{(l-1)} - i \tau \mathcal{E} ] e^{ i \tau (\Delta  + g_0 )}, \\
   \alpha_{2k  }^{(l)} &= [ \alpha_k^{(l-1)} - i \tau \mathcal{E} ] e^{ i \tau (\Delta - g_0  )},
  \end{array}
 \right.
\end{equation}
for $l \geq 2$ with the initial condition given by Eq.~(\ref{eq:alpha_1}). For odd $k$, coefficients $c$
are calculated as
\begin{equation}
\label{eq:coeff_l_o}
  \left\{
   \begin{array}{ll}
    c_{2k-1}^{(l)} &= ( \frac{1}{\sqrt{2}} )^l c_{k}^{(l-1)} e^{ -i \tau \mathcal{E} \mathrm{Re}
                         [  {\alpha_k^{(l-1)}} ] }, \\
    c_{2k  }^{(l)} &= ( \frac{-i}{\sqrt{2}} )^l c_{k}^{(l-1)} e^{ -i \tau \mathcal{E} \mathrm{Re}
                         [ {\alpha_k^{(l-1)}}] },
   \end{array}
  \right.
\end{equation}
and for even $k$ as
\begin{equation}
 \label{eq:coeff_l_e}
 \left\{
  \begin{array}{ll}
   c_{2k-1}^{(l)} &= ( \frac{-i}{\sqrt{2}} )^l c_{k}^{(l-1)} e^{ -i \tau \mathcal{E} \mathrm{Re}
                       [  {\alpha_k^{(l-1)}} ] }, \\
   c_{2k  }^{(l)} &=  ( \frac{1}{\sqrt{2}} )^l  c_{k}^{(l-1)} e^{-i \tau \mathcal{E} \mathrm{Re}
                       [  {\alpha_k^{(l-1)}} ] },
  \end{array}
 \right.
\end{equation}
for $l \geq 2$ with the initial condition given by Eqs.~(\ref{eq:alpha_1}) and (\ref{eq:coeff_1}).

The characteristics of the quantum walk dynamics are revealed when looking at the phase probability distribution of
state (\ref{eq:state_at_step_l}) in the phase space.
The coherent state $| \alpha_k \rangle$ rotates by an angle $\pm \Delta \theta = \pm \tau g_0$
in each time step [see Eq.~(\ref{shift_op})] which leads to a set of
discrete angles in the phase space [see~Fig.~\ref{fig1}(b)]. The Hilbert space for the quantum walk is
\begin{equation}
\label{eq:discrete_hilbert_space}
  \mathrm{H}_d = \mathrm{span}\biggl\{ | \varphi_m = \frac{2 \pi m}{d} \rangle =
                                        \frac{1}{\sqrt{d}} \sum_{j=0}^{d-1}
                                        e^{i j \varphi_m} | j \rangle;  m=0,1,\ldots,d-1 \biggr\},
\end{equation}
where $| \varphi_m \rangle$ are known as the phase states~\cite{loudon2000quantum} and play the role of the computational
basis. Here, $d$ is the Hilbert space dimension which is set to~$2 \pi / (\tau g_0)$.
The projection of the system state~(\ref{eq:state_at_step_l}) into $H_d$ is given by
\begin{align}
 \label{eq:state_at_step_l_projected}
 | \psi_l \rangle_d = \sum_{k=1}^{2^{l-1}} \sum_{m=1}^{d-1} \biggl\{
             &c_{2k-1}^{(l)} \langle \varphi_m | \alpha_{2k-1}^{(l)} \rangle | + \rangle | \varphi_m \rangle  \\ \nonumber
             &+c_{2k}^{(l)}  \langle \varphi_m | \alpha_{2k}^{(l)}    \rangle| - \rangle | \varphi_m \rangle 
             \biggr\}.
\end{align}
All coherent states in Eq.~(\ref{eq:state_at_step_l_projected}) are truncated when projected into~$\mathrm{H}_d$,
because $d$ is finite. However, if $d > |\alpha|^2 + |\alpha|$, where $|\alpha|^2$ represents the average number of photons
in the coherent state
and $|\alpha|$ is the corresponding standard deviation, the projection of the coherent state $|\alpha \rangle$ onto
Hilbert space $\mathrm{H}_d$ will be an acceptable approximation for $|\alpha \rangle$~\cite{sanders2003quantum}.
Moreover, since the coherent states have a minimum uncertainty
of  $1/2$~\cite{loudon2000quantum}, the value of $d$ cannot be arbitrarily large.
In fact, in a circle of radius $|\alpha|$ at most  $4\pi|\alpha|$ distinguishable
coherent states can be fitted. Therefore, the condition $d\le 4\pi|\alpha|$ must be imposed.
By equating the lower and the upper bounds for $d$ we obtain that $|\alpha|= 4\pi-1$. 
We conclude that the upper bound for the number of sites in the circle is around 145 
(a cycle with 145 vertices).

Using Eq.~(\ref{eq:state_at_step_l_projected}), the (unnormalized) probability distribution of the walker in the phase
space is calculated as
\begin{align}
 \label{eq:phase_prob_dist}
 P^{(l)}(m) = \biggl| \biggl(
                      \langle+|  \langle \varphi_m| \biggr) | \psi_l \rangle \biggr|^2 +
                      \biggl| \biggl( \langle-|  \langle \varphi_m| \biggr) | \psi_l \rangle \biggr|^2,
\end{align}
where the expression
\begin{equation}
 \label{eq:phase_states_coeff}
 \langle \varphi_m | \alpha_k^{(l)} \rangle = \frac{1}{\sqrt{d}} e^{-\frac{1}{2} \left|\alpha_k^{(l)}\right|^2}
  \sum_{j=0}^{d-1} \frac{ \left(\alpha_k^{(l)} e^{-i \varphi_m}\right)^j }{\sqrt{j!}},
\end{equation}
for $k = 1,\ldots 2^{l}$ is used in order to calculate $ \langle \varphi_m | \psi_l \rangle $.
In the next section it is discussed how to normalize $P^{(l)}(m)$.

\begin{figure}
\includegraphics[trim = 15mm 65mm 20mm 65mm, clip=true, width=8.6cm]{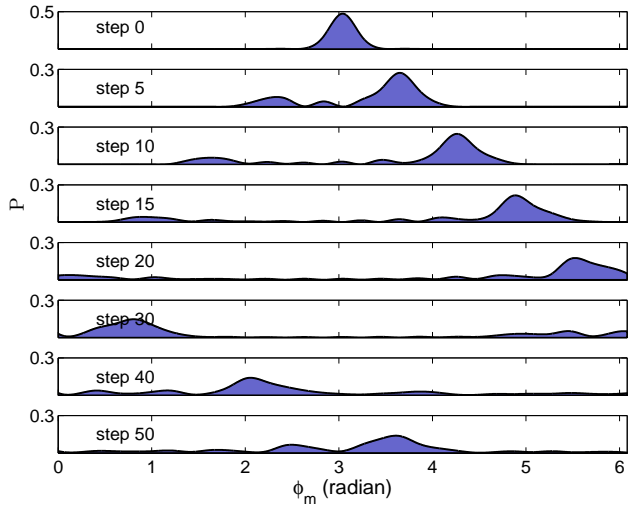}
\caption{(Color Online) The (unnormalized) phase probability distribution of the quantum walk for some time steps. Here, the initial state~(\ref{eq:asym_ini}) with $|\alpha_0|=5$ has been used, the
dimension of the Hilbert space has been set to $d=31$, and $\Delta=0$ and $\tau\mathcal{E}=0.001$.}
\label{fig:prob_dist_asym}
\end{figure}

Figure~\ref{fig:prob_dist_asym} shows $P^{(l)}(m)$ for some values of $l$ taking~(\ref{eq:asym_ini})
as initial state. Step $l=0$ corresponds to the initial distribution. The figure helps to see some characteristics of
the quantum walk propagation, which are remarkably different from
classical random walks. The probability distribution of quantum walks can have high peaks far from the origin, while
random walks are described by normal distributions.

\begin{figure}
\includegraphics[trim = 15mm 68mm 20mm 65mm, clip=true, width=8.6cm]{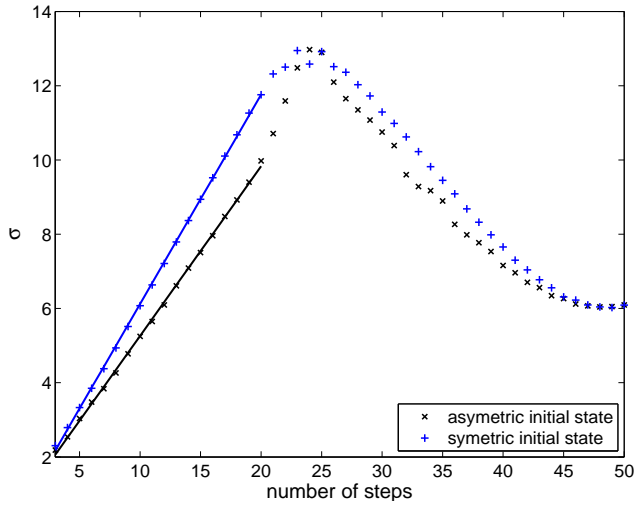}
\caption{(Color Online) The phase standard deviation as function of the number of steps for two initial states.
The black line ($\times$ points) uses state~(\ref{eq:asym_ini}) and the
blue line ($+$ points) uses state~(\ref{eq:new_sym_ini})
[compare with Figs.~\ref{fig:prob_dist_asym}~and~\ref{fig:prob_dist}].}
\label{fig:sigma_15steps}
\end{figure}

To further stress the difference between quantum walks and random walks, we analyze the dependence of 
the phase standard deviation ($\sigma$) as a function of time steps ($l$).
It is well known that $\sigma \propto \sqrt{l}$ for random walks. Figure~\ref{fig:sigma_15steps} shows $\sigma$ as
a function of $l$ for two initial states obtained from our simulations. Clearly, there is a linear dependence or a
ballistic behavior for both initial states. The black line ($\times$ points) can be approximately expressed
as {\color{black}$\sigma\approx 0.46\,l + 0.70$}, which uses initial state~(\ref{eq:asym_ini}).
The blue line ($+$ points) can be approximately expressed
as {\color{black}$\sigma\approx 0.56\,l + 0.49$}, which uses initial state~(\ref{eq:new_sym_ini}).
The linear dependence of the standard deviation with the time steps is a remarkable characteristic of the quantum walk
which may be checked in actual physical implementations.

{\color{black} Given the natural periodic boundary conditions of the quantum walk on a circle, it is interesting to check the typical effects expected on the probability distribution and standard deviation for time steps large enough to cover the entire circle. 
Figs.~\ref{fig:prob_dist_asym} and~\ref{fig:prob_dist} display a tendency of phase revival leading to a decrease of the standard deviation as shown
in Fig.~\ref{fig:sigma_15steps} for both the symmetric and asymmetric initial states. Notice that the characteristic linear dispersion occurs only until approximately 20 time steps and after that the standard deviation oscillates 
around some average value.}

{\color{black} It should be emphasized that our implementation proposal and 
similar proposals in the Literature, which encode the position 
space in the radiation-field phase space, are naturally suitable for implementing quantum walks on a circle. The existing experiments are mainly suitable for implementing quantum walks on a (theoretically infinite) line
because the realization of periodic boundary conditions is very difficult 
in principle. In our proposal it is possible to evolve the quantum walk with more time steps compared to previous experimental realizations.}

\subsection{Normalizing the probability distribution}
The areas under the plots depicted in Fig.~\ref{fig:prob_dist_asym} are not equal to one. 
This follows from the fact that state~(\ref{eq:state_at_step_l_projected}) has norm strictly smaller than one 
when $d$ is finite. The tails of the series of the coherent
states are lost when they are projected into Hilbert space $H_d$.
Moreover, the area under the plot is changing during the evolution of the system, because there is a vertical
displacement of the coherent state in the phase space produced by operator
$e^{- i \tau \mathcal{E} (a^{\dagger} + a)}$,
affecting $\alpha$'s
[Eq.~(\ref{eq:alpha's})] in each time step
and consequently modifying 
the norm of $| \psi_l \rangle_d$.

For sufficiently small $\tau \mathcal{E}$, the norm of $| \psi_l \rangle_d$ does not depend on $l$
and is almost equal to the norm of the truncated initial coherent state
\begin{equation}
\label{eq:truncated_coherent_state_norm}
  \langle \alpha_0 | \alpha_0 \rangle_d =
   e^{-|\alpha|^2} \sum_{j=0}^{d-1} \frac{|\alpha|^{2j}}{j!}.
\end{equation}
By setting $\mathcal{E}=0$ and $\Delta=0$ in the propagator~[Eq.~(\ref{eq:U})],
a standard quantum walk on a cycle would be obtained.
To normalize $P^{(l)}(m)$, $| \psi_0 \rangle_d$ should be normalized by dividing it by
$\sqrt{ \langle \alpha_0 | \alpha_0 \rangle_d } $.

The effect of nonzero $\Delta$ when $\tau \mathcal{E}$ is sufficiently small is a free shift by
angle $e^{i \tau \Delta}$ in each time step independently of the coin state. This term introduces
a bias which can be clockwise or counterclockwise depending on the sign of $\Delta$.

For large $\tau\mathcal{E}$ the position of the walker goes off the cycle destroying the ballistic dynamics of the walk.
Therefore it is important to keep it small.
Since $\tau =\pi/2\Omega_m$ and $\Omega_m$ ranges from KHz to GHz \cite{aspelmeyer2013cavity},
it implies that $\mathcal{E}$ can be sufficiently large and still the condition 
$\tau\mathcal{E} \ll 1$ is satisfied. 
{\color{black}Numerical simulations show the effect of $\mathcal{E}$ is negligible when $\tau\mathcal{E}$ is less
than $10^{-3}$. Considering the range of $\tau$, the value of $\mathcal{E}$ ranges from $1$ to $10^6$ Hz.}

\subsection{Symmetric probability distribution}
The Hadamard-like transformation (\ref{eq:hadamard}) treats coin states $|+\rangle$ and $|-\rangle$ in the same way and
does not bias the walk~\cite{Kempe:03b}. Therefore starting the walk with the symmetric state
$ (1/\sqrt{2}) ( |+\rangle + |-\rangle ) | \varphi_m \rangle$
leads to a symmetric distribution around $\varphi_m$.
Starting the walk in a superposition of states $ | \varphi_m \rangle$ (non-local state), so that each of them is symmetric,
still produces a symmetric probability distribution if the amplitudes are real.
However the truncated coherent state $| \alpha_0 \rangle_d$ is a superposition of phase states with complex amplitudes.
Distinct phase factors will interfere during the evolution destroying the symmetry.

\begin{figure}
\includegraphics[trim = 15mm 68mm 20mm 65mm, clip=true, width=8.6cm]{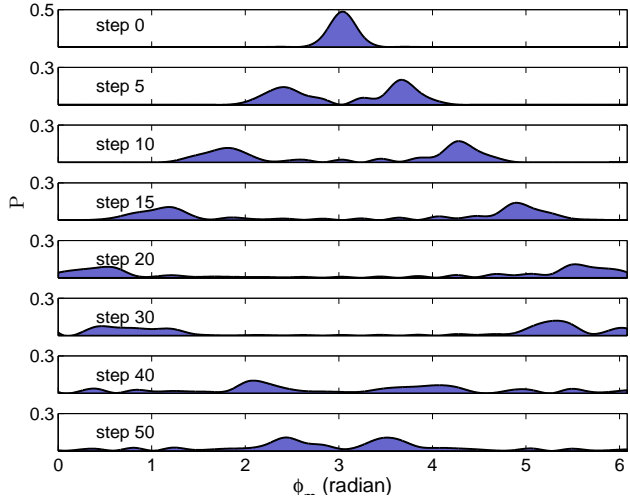}
\caption{(Color Online) The (unnormalized) phase probability distribution of the quantum walker for some time steps. Here, the initial state~(\ref{eq:new_sym_ini}) with $|\alpha_0|=5$ has been used, the
dimension of the Hilbert space has been set to $d=31$, $\Delta=0$, and $\tau\mathcal{E}=0.001$.}
\label{fig:prob_dist}
\end{figure}

The (unnormalized) probability distributions with the initial state
\begin{equation}
 \label{eq:new_sym_ini}
 | \psi_0' \rangle = \frac{1}{\sqrt{2}} \biggl( | + \rangle + | - \rangle \biggr) | \alpha_0 \rangle
\end{equation}
are depicted in Fig.~\ref{fig:prob_dist} for some values of $l$.
It must be mentioned that with the initial state~(\ref{eq:new_sym_ini}),
Eqs.~(\ref{eq:coeff_1}) is slightly modified. In this case, the
walker evolves in a more symmetric way than the case with the initial state~(\ref{eq:asym_ini}).

\section{\label{sec:decoherence}Decoherence}
Both the optical and the mechanical resonator can be affected by noise.
As we have discussed, the dissipation effect on the optical field is counteracted by the driving so that it is left in an
undepleted coherent state of amplitude $|\alpha|\propto {\mathcal{E}}/{\gamma}$,
where $\gamma$ is the relaxation constant for the amplitude damping channel (see for example \cite{marcos1}).
However, the noise effects are more severe over the mechanical resonator, particularly in the situation where only
two lowest levels are relevant. Therefore, the coherence time of this two-level resonator limits the total random
walk evolution time. We consider the effect of the dephasing channel on the two-level resonator, or in terms
of the quantum walk interpretation, we analyze a decoherent coin.

The unitary evolution of the quantum walk in each time step is driven by
the unitary operator $U$ given by Eq.~(\ref{eq:U}).
In the presence of imperfections the quantum walk evolution is not unitary and can be described by
\begin{equation}
 \rho_l = \sum_{j} K_j U \rho_{l-1} U^{\dagger}  K_j^{\dagger},
\end{equation}
where $\rho_l$ is the density operator of the walker plus coin system and $K_j$ are the Kraus operators modeling the
quantum noise.
The effect of the phase damping channel on the two-sided coin can be modeled by the Kraus
operators $K_j= I_w\otimes E_j, \; j=0,1$ where $I_w$ is the identity operator on the walker space
and~\cite{nielsen2010quantum}
\begin{equation}
\label{eq:phase_damp}
 E_0 = \begin{pmatrix}
        1& \;\; 0 \\
        0& \;\; \sqrt{1-\lambda} \\
       \end{pmatrix},\;\;\;\;\;
 E_1 = \begin{pmatrix}
        0& \;\; 0 \\
        0& \;\; \sqrt{\lambda} \\
        \end{pmatrix},
\end{equation}
where $\lambda$ quantifies the strength of the channel and can be written as $\lambda=1 - e^{-l\tau/T_d}$,
being $T_d$  the dephasing time.
The amount of phase damping increases as time goes on.

Using the results of the previous section,
the density operator of the system at step $l$ is written as
\begin{align}
 \rho_l = \sum_{k,k'}
           \bigl\{ \;
            r_{2k-1,2k'-1}^{(l)}  \; &|+\rangle\langle+| \; | \alpha_{2k-1} \rangle \langle \alpha_{2k'-1} | \\ \nonumber
            +r_{2k-1,2k'  }^{(l)} \; &|+\rangle\langle-| \; | \alpha_{2k-1} \rangle \langle \alpha_{2k'  } | \\ \nonumber
            +r_{2k  ,2k'-1}^{(l)} \; &|-\rangle\langle+| \; | \alpha_{2k  } \rangle \langle \alpha_{2k'-1} | \\ \nonumber
            +r_{2k  ,  2k'}^{(l)} \; &|-\rangle\langle-| \; | \alpha_{2k  } \rangle \langle \alpha_{2k'  } |
           \;\bigr\},
\end{align}
where $1 \leq l \leq n$ and the coefficients $r$ for $l\geq2$ are given by
\begin{equation}
\left\{
        \begin{array}{llll}
            r_{2k-1,2k'-1}^{(l)} &= \frac{1}{2} \;  r_{k,k'}^{(l-1)}  \\
            r_{2k-1,2k'  }^{(l)} &= -\frac{ i}{2} (-1)^k \; r_{k,k'}^{(l-1)}  \\
            r_{2k  ,2k'-1}^{(l)} &= \frac{ i}{2}  (-1)^k \; r_{k,k'}^{(l-1)}  \\
            r_{2k  ,  2k'}^{(l)} &= \frac{1}{2} \;  r_{k,k'}^{(l-1)},
        \end{array}
\right.
\end{equation}
when $k$ and $k'$ have the same parity and 
\begin{equation}
\left\{
        \begin{array}{llll}
            r_{2k-1,2k'-1}^{(l)} &= - \frac{i}{2}(-1)^k\sqrt{1-\lambda} \;  r_{k,k'}^{(l-1)}  \\
            r_{2k-1,2k'  }^{(l)} &= \frac{1}{2}\sqrt{1-\lambda} \;  r_{k,k'}^{(l-1)}  \\
            r_{2k  ,2k'-1}^{(l)} &= \frac{1}{2}\sqrt{1-\lambda} \;  r_{k,k'}^{(l-1)}  \\
            r_{2k  ,  2k'}^{(l)} &= \frac{ i}{2}(-1)^k\sqrt{1-\lambda} \; r_{k,k'}^{(l-1)},
        \end{array}
\right.
\end{equation}
when $k$ and $k'$ have different parities. The initial conditions for the symmetric initial coin state
are $r_{1,1}^{(1)}= r_{2,2}^{(1)} = 1/2$, and $ r_{1,2}^{(1)}=r_{2,1}^{(1)}=\sqrt{1-\lambda}/2$.

The (unnormalized) probability distribution in the phase space is then calculated as
\begin{align}
 \label{eq:phase_prob_dist_density_op}
 P^{(l)}(m) &= \langle+|  \langle \varphi_m| \rho_l | \varphi_m \rangle |+\rangle \\ \nonumber
            &+ \langle-|  \langle \varphi_m| \rho_l | \varphi_m \rangle |-\rangle \\ \nonumber
            &= \sum_{k,k'} r_{2k-1,2k'-1}^{(l)} \langle \alpha_{2k-1} | \varphi_m \rangle
                                                        \langle \varphi_m | \alpha_{2k'-1} \rangle \\ \nonumber
            &+ \sum_{k,k'} r_{2k  ,2k'  }^{(l)} \langle \alpha_{2k} | \varphi_m \rangle
                                                        \langle \varphi_m | \alpha_{2k'} \rangle \bigr\}.
\end{align}

\begin{figure}
\includegraphics[trim = 15mm 68mm 20mm 65mm, clip=true, width=8.6cm]{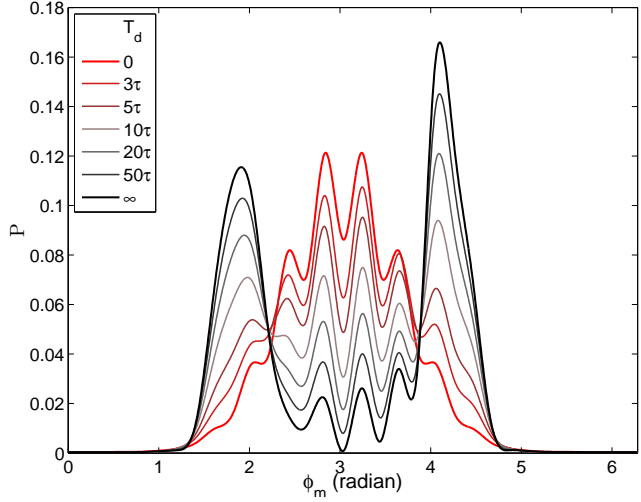}
\caption{(Color Online) The (unnormalized) phase probability distribution of the quantum walker after $l=9$ steps
for different values of dephasing time $T_d$.
Here, the initial state~(\ref{eq:new_sym_ini}) has been used.}
\label{fig:decoherence}
\end{figure}

Fig. ~\ref{fig:decoherence} depicts $P^{(l)}(m)$ after $l=9$ time steps for different values of dephasing time $T_d$.
The figure shows the quantum-to-classical transition of the quantum walk dynamics when dephasing time decreases.

\begin{figure}
\includegraphics[trim = 15mm 68mm 20mm 65mm, clip=true, width=8.6cm]{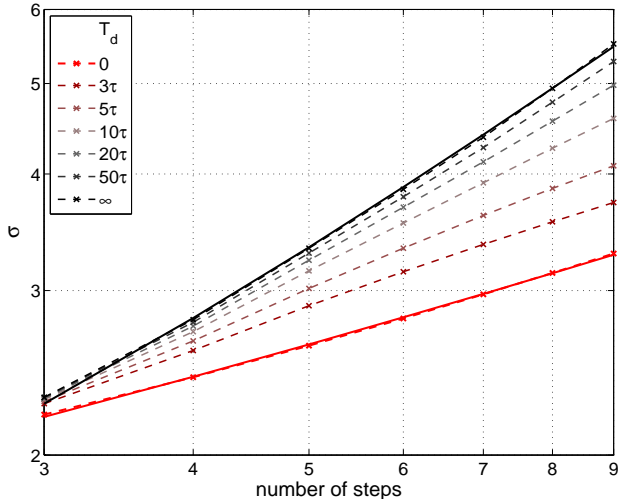}
\caption{(Color Online) The standard deviation of the walker in terms of the number of steps $l$ for different values of
dephasing time $T_d$  in log-log scale [compare with Fig.~\ref{fig:decoherence}].}
\label{fig:sigma}
\end{figure}

Fig.~\ref{fig:sigma} shows the phase standard deviation $\sigma$ as a function of time steps $l$
for different values of dephasing time $T_d$ in the log-log scale.
The black continuous line shows the linear
dependence of $\sigma$ with the number of steps in the perfect case $T_d=\infty$.
The linear dependence of $\sigma$ gradually converts to the curve $\sqrt l$  when the dephasing time decreases.
The red continuous curve is $\sigma\propto\sqrt{l}$ and indicates that $\sigma$
evolves proportional to the square root of time steps when decoherence is maximum.
We see therefore a rapid degrading of the quantum feature on the walk for a dephasing time $T_d$ smaller than
tens of the time step $\tau$. Collecting the parameters necessary for this implementation we know that on
typical optomechanical experiments \cite{aspelmeyer2013cavity} the mechanical resonator
frequency $\Omega_m$ ranges from KHz to GHz and therefore the time step $\tau$ ranges from $10^{-3}$ to $10^{-9}$ seconds.
Depending on the specific implementation the optical and mechanical relaxation rate may vary enormously.
Particularly in the electromechanical system discussed in \cite{Teufel2009}
$\Omega_m \approx 10^{7}$ Hz and $1/T_d\approx 10^{1}$ Hz and therefore $T_d\approx 10^6 \tau$ implying that
the decoherence on the coin is almost negligible and so the quantum walk can be implemented over a large
number of steps $n$. Similar conclusions are valid for the optomechanical system in \cite{Verhagen}.

{\color{black} Our decoherence analysis is performed in the coin rotated basis in which the quantum walk is described. The decoherence model was chosen
based on the fact that the worst noise effect is a bit-flip channel in the coin original basis, which is produced by uncontrolled transitions between the ground and excited states. This corresponds to the phase-flip channel in the rotated basis, which is equivalent to the phase damping due to the unitary freedom. Notice that if we had considered a phase-damping channel in the original basis, corresponding to a bit-flip channel in the coin rotated basis, the effects would have been similar to the ones described in Ref.~\cite{chandrashekar2007symmetries}.}
\section{Discussions and Conclusions}\label{sec:discussion}

In order to implement the discrete-time quantum walk model, two separate unitary operators are required, namely, the coin and
the shift operators. The corresponding Hamiltonians do not commute
and, by alternating their action, the quantum walk dynamics is generated. Usually
the shift Hamiltonian is kept fix and a sequence of pulses are applied with a defined duration and frequency to implement
the coin operator from time to time. The application of such pulse sequence, however, introduces noises and disturbs the walker's
position.

In Sec.~\ref{sec:QW} we showed how the time evolution of Hamiltonian~(\ref{eq:standard_om_hamiltonian_two_level_rotated})
can simulate the quantum walk dynamics, when the parameters are properly adjusted.
The distinct feature of the described method is that the quantum walk dynamics is obtained by the natural evolution of
Hamiltonian~(\ref{eq:standard_om_hamiltonian_two_level_rotated}) with no need to switch between the individual Hamiltonians
or applying any pulse sequence.
The success of that method is due to the fact that for sufficiently small time intervals it is possible to decompose 
the time evolution operator as a product of operators corresponding to the individual coin and shift Hamiltonians. 

In order to obtain the ideal quantum walk dynamics, we have restricted the system parameters {\color{black} during our
calculations.} Such restrictions seem artificial and impossible to impose on the natural evolution of the system, and seem to
lead to a large approximation error. However, in the natural evolution of the system we can {\color{black}play with the
system frequencies $g_0$ and $\Omega_m$. A possible procedure to set up the system is as follows.
In typical optomechanical experiments as mentioned at the end of Sec.~\ref{sec:decoherence},  the mechanical
resonator frequency $\Omega_m$ ranges from  $10^{3}$ to $10^{9}$ Hz. The time step $\tau$ can be fixed by choosing a
frequency and by using $\tau=\pi/2\Omega_m$ [Eq.~(\ref{eq:coin_cond})], which is between $10^{-3}$ and $10^{-9}$ seconds.
For the moment, the interaction strength $g_0$ is set in such way that it obeys $g_0 \tau = 2 \pi / d$
[below Eq.~(\ref{eq:discrete_hilbert_space})], for an appropriate dimension $d$.
However, with such parameters the resulting dynamics in the natural evolution of the system simulates a quantum walk with a
coin very close to the identity. To obtain a quantum walk with the Hadamard coin we need to decrease the ratio $g_0/\Omega_m$.
Numerical simulations shows that by decreasing that ratio by a factor around $10$, the Hadamard coin is realized.
}


We have also considered a minimum uncertainty of $1/2$ for the coherent state
[below Eq.~(\ref{eq:state_at_step_l_projected})] which is a better lower bound when $|\alpha| \geq 1$.
In this way, we obtain the upper bound $4\pi|\alpha|$ for the maximum number of sites which is twice the value
suggested in Ref.~\cite{sanders2003quantum}.


To sum up, we have presented a proposal for implementation of a quantum walk on a circle using optomechanical systems.
The walker is described by the coherent state in the phase space of the light field while the coin is encoded on
the states of a mechanical resonator.  We have illustrated the process by considering the simplest case of a
two-sided coin where only the two-lowest states of the mechanical resonator are relevant. The coherent state moves
around a circle in the phase space so that the number of sites and its radius
can be chosen by tuning the number of photons in the optical cavity. The number of sites in the cycle is limited
by around 145. Despite that the total number of steps of the walker is not limited, but periodic in the cycle. 

We have performed numerical simulations of the dynamics of the proposed scheme, which display the signatures of the
discrete-time coined quantum walk on cycles. The ballistic behavior can be verified in two ways: (1)~the probability
distribution has two high peaks moving in opposite directions away from the origin, and (2)~the standard deviation
of the phase probability distribution is proportional to the number of steps.   It is well known that classical
random walks have a normal probability distribution and the standard deviation is proportional to the square
root of the number of steps. 

We have analyzed the decoherence effects over the coin by using the phase damping channel. We conclude that the decoherence
time is large enough to allow the observation of the quantum behavior with present-day experimental parameters.
As a last remark we should mention that for practical purposes the phase of the walker is easily detected
through standard homodyne measurements, while  the measurement of the population of the two lowest states of the mechanical
resonator
requires a more involved procedure. This is currently a subject of intense research and the details for that detection
depends on the specific optomechanical or electromechanical implementation. In fact
the experiments recently reported in Refs. \cite{okamoto2013coherent,faust2013coherent} open a new and promising perspective.

We are currently extending the quantum walk model to analyze localization and to implement spatial search algorithms
on two-dimensional lattices, by using two coupled optomechanical resonators---an aspect which shall be addressed elsewhere.
We believe that physical realizations of quantum walks on optomechanical systems may bring practical
applications quicker than the realization of a general-purpose quantum computer.

\begin{acknowledgements}
JKM acknowledges financial supports from CNPq, grants PCI-DB 302866/2014-0 and PDJ 165941/2014-6.
RP acknowledges financial support from CNPq and FAPERJ. MCO acknowledges support by FAPESP and CNPq through the National Institute for Science and Technology of Quantum Information (INCT-IQ) and the Research Center in Optics and Photonics (CePOF). 
\end{acknowledgements}

\end{document}